\documentclass[aps,prx,reprint,showpacs,superscriptaddress,floatfix]{revtex4-2}
\usepackage[utf8]{inputenc}
\usepackage{graphicx}
\usepackage{booktabs}
\usepackage{tabularx}
\usepackage{hyperref}
\hypersetup{
colorlinks=true,final=true,
        linkcolor=blue,
        citecolor=blue,
        filecolor=blue,
        urlcolor=blue,
}

\begin{document}

\title{Electronic structure of higher-order layered palladates: {La$_{n+1}$}{Pd$_{n}$}{O$_{2n+2}$} $(n = 4-7)$}

\author{Alexander Gavrilov}
\email{agavrilo@asu.edu}
\affiliation{Department of Physics, Arizona State University, Tempe, AZ - 85287, USA}

\author{Lidia C. Santander}
\affiliation{Department of Physics, Arizona State University, Tempe, AZ - 85287, USA}

\author{Yifeng Zhao}
\affiliation{Department of Physics, Arizona State University, Tempe, AZ - 85287, USA}

\author{Antia S. Botana}
\affiliation{Department of Physics, Arizona State University, Tempe, AZ - 85287, USA}

\date{\today}

\begin{abstract}

The square-planar layered nickelates R$_{n+1}$Ni$_n$O$_{2n+2}$ (R = Nd, $n=4-7$) have been recently shown to be superconducting without the need for chemical doping or pressure. Here, we examine the electronic structure of the analog higher-order square-planar palladates --that have not yet been synthesized-- via \textit{ab initio} calculations. These layered palladates exhibit larger bandwidths, an increased $p-d$ hybridization, less interference from R-$d$ bands at the Fermi level, and antiferromagnetic ground states. These characteristics make them closer cuprate analogs and promising candidates to pursue in the context of unconventional superconductivity. 

\end{abstract}

\maketitle{}

\section{Introduction} \label{intro}

One strategy to solving the long-standing problem of high-temperature superconductivity in the cuprates has been to look for materials with a similar structure and $3d$ electron count \cite{Norman_2016}. Using this approach, much progress has been made by analyzing nickelates as cuprate analogs \cite{Anisimov1999-at, Lee2004-mh, mitchell_review, Puphal_review}. An important breakthrough took place in 2019 when superconductivity (with $T_c$ $\sim$ 15 K) was discovered in the so-called infinite-layer nickelate NdNiO$_2$ upon hole doping \cite{Li2019}. In the infinite-layer nickelates, the Ni atoms realize a hard-to-stabilize Ni$^{1+}$ oxidation state corresponding to a $d^9$ electron count, analog to Cu$^{2+}$. Furthermore, their structure portrays infinite NiO$_2$ planes, akin to the CuO$_2$ planes of the cuprates. Importantly, RNiO$_2$ (R= rare-earth) materials are simply one member of a larger structural family of square-planar layered nickelates with general chemical formula R$_{n+1}$Ni$_n$O$_{2n+2}$, where $n$ represents the number of NiO$_2$ planes along the $c$ axis \cite{LACORRE1992495, Greenblatt1997-xo}. While the Nd-based quintuple layer ($n=5$) member of the family has been known to be superconducting since 2022 \cite{Pan_Ferenc}, it has only recently been shown that a superconducting dome extends from the $n=4$ to the $n=7$ members of the series, with similar maximum $T_c$ $\sim$ 15 K \cite{Pan2026}.  

In spite of the similarities between square-planar layered nickelates and cuprates in terms of their structure and 3$d$ electron count, they exhibit several important differences in terms of their electronic structure and magnetism \cite{Kitatani2020, Nica2020, Botana2020, LaBollita2021, Zhang2020-zd, Petocchi2020-gf, Choi2020-an, Lechermann2020-mj, Lechermann2020-na, Kapeghian2020-wo, Leonov2020-zi, Karp2020-yz, Gu2020, Ryee2020-fc, arita, ES_112, Hu2019-ep,
ES_112, arita, Liu2020, Thomale_PRB2020, Choi2020, Ryee2020, Gu2020,   Sakakibara, jiang2020, werner2020, eff_ham, Zhang2020, Wang2020}. The degree of $p-d$ hybridization is much weaker in the nickelates and they do not display single-band physics, but instead have extra bands of R-$d$ character at the Fermi level, unlike cuprates. Furthermore, there is no evidence for the presence of long-range magnetic order at $d^9$ filling in the square-planar nickelates, even though the existence of antiferromagnetic correlations has been experimentally confirmed \cite{jexp_nickelates} (with a lower superexchange $\sim$ 60 K). These relevant differences may be linked to the lower T$_c$ that the reduced nickelates display when compared to cuprates.  Hence, analyzing other transition metal oxides with a similar square-planar arrangement and $3d$ electron count can be one strategy to pinpoint the relevance of several cuprate characteristics for superconductivity and, if realized, to understand how those may scale with T$_c$.

\begin{figure}
    \centering
    \includegraphics[width=\columnwidth]{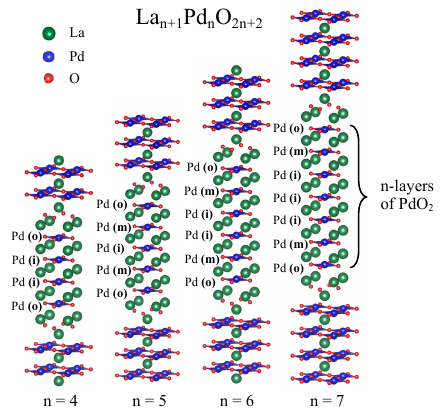}
    \caption{Crystal structures of the square-planar layered palladates with $n$ representing the number of PdO$_2$ layers along the $c$ axis. Also labeled is the naming convention for inner(i), middle(m), and outer(o) layers for each of the structures.}
    \label{fig1}
\end{figure}

In the vein of finding layered-nickelate analogs, palladates are an obvious target as Pd is below Ni in the periodic table \cite{kitatani_palladates, Botana2018}. Indeed, work using dynamical vertex approximation proposed that palladates such as RbSr$_2$PdO$_3$ and BaLaPdO$_2$Cl$_2$ may be promising superconducting candidates with higher T$_c$s than the layered nickelates \cite{kitatani_palladates}. In light of the recent discovery of superconductivity in Nd$_{n+1}$Ni$_n$O$_{2n+2}$ ($n= 4-7$), it is worth analyzing the electronic structure of the analog Pd compounds, that have not yet been experimentally realized. We note that in contrast to the layered nickelates (that need to be reduced from parent perovskite or Ruddlesden-Popper phases \cite{Li2019, Pan2026}) it may be possible to synthesize the palladates directly in square-planar format  due to the stability of Pd$^{1+}$ in contrast to Ni$^{1+}$. 

Here, we explore the crystal and electronic structure properties of the hypothetical higher-order layered palladates R$_{n+1}$Pd$_n$O$_{2n+2}$ (R = La, $n= 4-7$), whose structures are shown in Fig. \ref{fig1}. Our results reveal that the electronic structure of these palladate compounds lies between cuprates and nickelates-- with a Fermi surface topology closer to the single $d_{x^2-y^2}$-band picture and a larger degree of $p-d$ hybridization. As such, these palladates may reveal unexplored pathways towards identifying new superconducting materials.

\section{Methodology} \label{DFT}
The electronic structure calculations for square-planar palladates were done using the all-electron, full potential WIEN2K \cite{Blaha2020} code. For the exchange correlation functional, the Perdew-Burke-Ernzerhof version of the generalized gradient approximation (GGA-PBE) was used \cite{gga_pbe}. 
La was chosen as the rare-earth element to avoid the complexities that would come with 4$f$ electrons in other rare-earth cations.  Muffin-tin (MT) radii of  2.43, 2.10, and 1.81 a.u. were used for La, Pd, and O, respectively. An R$_{\mathrm{MT}}$K$_{\mathrm{max}}$ = 6.0 was employed. A dense $k$-mesh of up to $36\times36\times12$ was required to reach convergence. In order to gain further insights into the electronic structure,  maximally localized Wannier functions (MLWFs) were constructed from the DFT results in the nonmagnetic state. For the construction of the MLWFs, WANNIER90 \cite{Mostofi2014} and WIEN2WANNIER \cite{Kunes2010} were used. A wide energy window was employed to include the Pd(4$d$), O(2$p$), and La(5$d$) orbitals. To obtain a systematic comparison with the square-planar layered nickelates, the results of Ref. \cite{LaBollita2021} were independently reproduced and validated.  
Through the use of spin-polarized calculations, we have also investigated the stability of different magnetic configurations by constructing $\sqrt{2}\times\sqrt{2}$ supercells for the $n = 4-7$ layered palladates. These magnetic calculations were performed using GGA+$U$. We employed the `fully localized limit' (FLL) as the double-counting scheme \cite{fll}. $U$ values of 2-4 eV were used and we have chosen a non-zero $J=0.7$ eV in our calculations to properly account for the anisotropy  of the interaction. 

\section{Structural Properties} \label{StructProp}

 The palladates we analyze here have the general chemical formula La$_{n+1}$Pd$_{n}$O$_{2n+2}$, analogous to the reduced square-planar layered nickelates La$_{n+1}$Ni$_{n}$O$_{2n+2}$. We focus on the $n=4-7$ cases, which correspond to La$_{5}$Pd$_{4}$O$_{10}$ ($n=4$), La$_{6}$Pd$_{5}$O$_{12}$ ($n=5$), La$_{7}$Pd$_{6}$O$_{14}$ ($n=6$) and La$_{8}$Pd$_{7}$O$_{16}$ ($n=7$). Each square-planar palladate (nickelate) contains $n$-PdO$_2$(NiO$_2$)
 planes that are separated along the $c$-axis by a fluorite-like LaO
blocking slab (see Fig. \ref{fig1}). This blocking layer suppresses the $c$-axis dispersion with respect to the infinite-layer counterpart. 
The reduced palladate structures were constructed from the analog square-planar nickelates: all of them are tetragonal, belonging to the $I4/mmm$ space group symmetry and were fully relaxed (both lattice constants and internal coordinates).  Relevant structural data after relaxation is provided in Table \ref{table1} comparing nickelates with their analog palladates. While the $a$ and $b$ lattice constants (identical for all $n$s) are larger in the palladates with respect to the corresponding nickelates, the $c$ lattice constant is reduced in the palladates. These trends are in agreement with previous work on the infinite-layer palladate \cite{kitatani_palladates}.

\begin{table}[htbp]
    \centering
    \caption{Lattice parameters for higher-order layered nickelates and palladates.  All values are given in \AA.}
    \vspace{0.25cm}
    \label{tab:lattice_parameters}
    \begin{tabular}{l@{\hskip 15pt}l@{\hskip 15pt}l@{\hskip 15pt}l@{\hskip 15pt}l}
        \hline
        \hline
        \textit{} & \textit{$n=4$} &
        \textit{$n=5$} & 
        \textit{$n=6$} & 
        \textit{$n=7$} 
        \\
        \hline
        \textit{$a_{\mathrm{Ni}} = b_{\mathrm{Ni}}$} & 3.96 & 3.97 & 3.96 & 3.97\\
        \textit{$a_{\mathrm{Pd}} = b_{\mathrm{Pd}}$} & 4.17 & 4.17 & 4.18 & 4.18\\
        \textit{$c_{\mathrm{Ni}}$} & 32.99 & 39.93 & 46.06 & 53.52 \\
        \textit{$c_{\mathrm{Pd}}$} & 32.52 & 39.20 & 45.87 & 52.50 \\
        \hline
        \hline
    \end{tabular}
    \label{table1}
\end{table}

\begin{figure*}
    \centering
    \includegraphics[width=\textwidth]{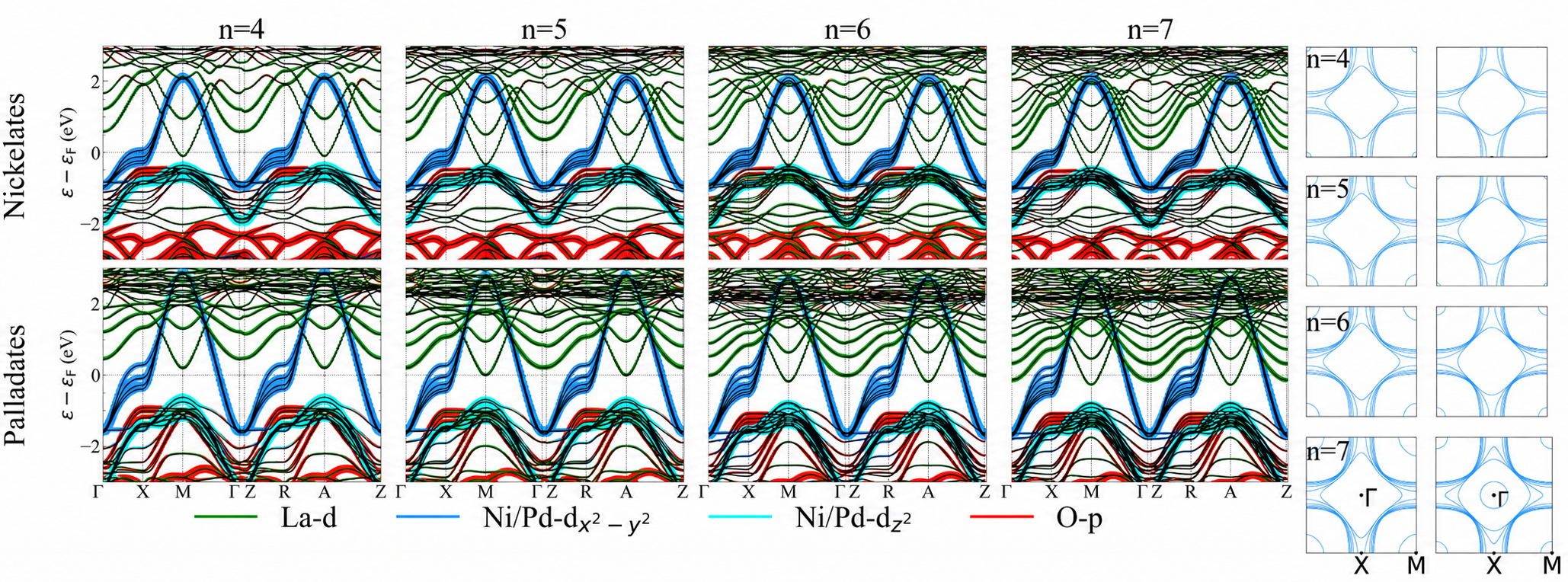}
    \caption{Comparison of the band structures with orbital character highlighted for the square-planar nickelates (top row) and palladates (bottom row.) Each column corresponds to the same number of layers in the materials, starting with $n = 4$ up to $n = 7$ layers. The right panels show the corresponding Fermi surface cuts at $k_z$=0.}
    \label{fig2}
\end{figure*}

  In both families the out-of-plane La-La distances are modulated through the $n$-layer blocks. The La-La distances are smaller in the inner layers, approaching the $c$-axis lattice constant of the infinite-layer material. Towards the edge of the $n$-layer block close to the fluorite slab (outer layers), the La-La distance increases in both nickelates and palladates by $\sim$ 0.1 \AA~ (from $\sim$ 3.37 to 3.47 \AA). Our finding of modulated R-R bond lengths is consistent with recent STEM data in the finite-$n$ nickelates \cite{Pan2026}. Further, in the nickelates, the reverse modulation takes place for the out-of-plane Ni-Ni distance that is larger in the inner layers, and smaller towards the outer layers. In contrast, in the palladates, the Pd-Pd out-of-plane distance does not significantly modulate across layers. Importantly, the Pd-Pd bond length is shorter than the average interlayer Ni-Ni distance ($\sim$3.26 vs. $\sim$3.30 \AA).  Overall, our results reflecting layer-modulated bond lengths suggest that other electronic properties may also be layer-dependent in this family of materials (see Section IV).

\section{Non-magnetic Electronic Structure} \label{eStruct}

\begin{figure}
    \centering
    \includegraphics[width=\columnwidth]{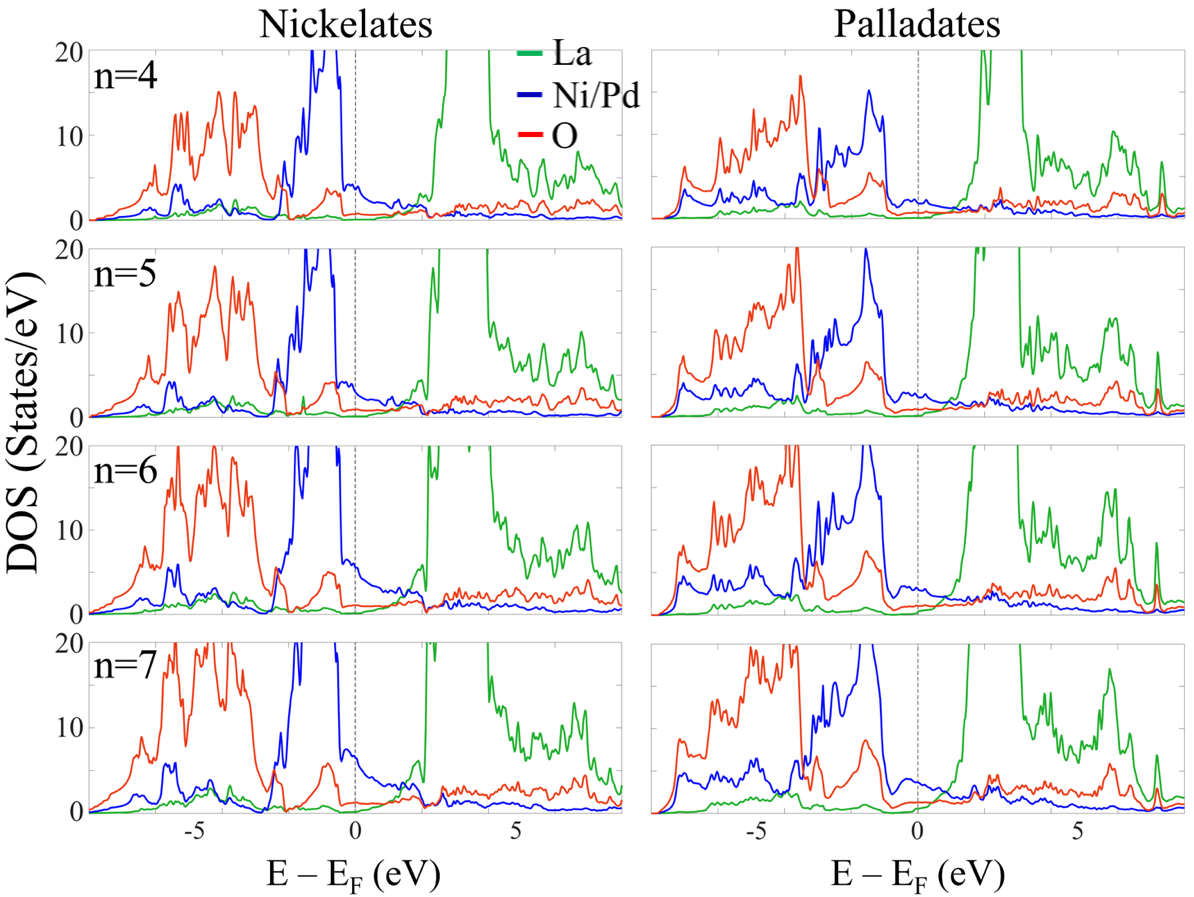}
    \caption{Comparison of the atom-resolved density of states for the square-planar nickelates (left column) and palladates (right column). Each row corresponds to the same number of layers in the corresponding materials analyzed, starting with $n = 4$ up to $n = 7$.}
    \label{fig3}
\end{figure}

The nominal Ni/Pd $d$-filling for materials in the square-planar layered nickelate/palladate families can be expressed as a function of the number of layers $n$ as $d^{9 - 1/n}$. The superconducting dome area of filling for the $n=4-7$ nickelates falls between ~$d^{8.75}$ to ~$d^{8.86}$, matching cuprate-like values \cite{Keimer_2015}. Figure \ref{fig2} shows the band structures of the higher-order layered nickelates in comparison to the analog palladates. For the nickelates (top panels),
$n$ bands of $d_{x^2 - y^2}$ character 
cross the Fermi level, akin to the cuprates. Unlike cuprates, however, extra bands of La-$d$ character also cross the Fermi level in all cases  ($n=4-7$) at the M and A points of the Brillouin zone, giving rise to electron pockets that create a self-doping effect for the $d_{x^2-y^2}$ bands. 
While the role that these bands play in the physics of reduced layered nickelates is still debated, several works \cite{LaBollita2022, Kitatani2020, Karp2020-yz} indicate  that they may simply act as a charge reservoir, similar to the role that O-$p$ states play in the cuprates. 
Figure \ref{fig3} (left panels) shows the corresponding atom-resolved DOS curves of the layered nickelates reflecting a large $p-d$ splitting when compared to cuprates, giving then rise to a larger charge-transfer energy ($\sim$ 4 eV as estimated from the on-site energies of wannierizations, in contrast to the $\sim$ 2 eV values obtained in the cuprates \cite{LaBollita2021}). 

We now contrast the electronic structure of the $n= 4-7$ layered nickelates with that of the corresponding palladates (as shown in the bottom panels of Fig. \ref{fig2}). Like in the nickelate case, $n$-Pd-$d_{x^2-y^2}$ bands can be seen to cross the Fermi level. These bands of $d_{x^2 - y^2}$ character have much larger bandwidth in the palladates when compared to the nickelates ($\sim$ 3  vs. 4.5 eV), linked to a larger degree of $p-d$ hybridization. Furthermore, the $d_{z^2}$ orbitals are shifted to considerably lower energies in the palladates. In contrast to the nickelates (where the La-$d$ bands already cross the Fermi level for the $n = 4$ compound), in the palladates the La-$d$ bands only start to cross the Fermi level in the $n = 6$ case, leading overall to 
less involvement of the La-$d$ states at the Fermi level in the Pd compounds. The corresponding Fermi surface cuts at $k_z$=0 (see Fig. \ref{fig2}) clearly show the reduced involvement of La-$d$ orbitals in the palladates, reflected in the absence of electron pockets in the corner of the zone (M) up to the quintuple-layer palladate, in contrast to the corresponding nickelates. The other ($n$) Fermi surface sheets are predominantly Pd/Ni-$d_{x^2-y^2}$ in character with one being electron-like (centered at $\Gamma$) and the rest hole-like (centered at M). Note that the $k_z$= 1/2 cut is identical, given the 2D character of the finite-$n$ compounds, due to the presence of the blocking fluorite slab. 
Overall, in the higher-order layered palladates, a more single-band cuprate-like picture is hence recovered.

\begin{figure}
    \includegraphics[width=\columnwidth]{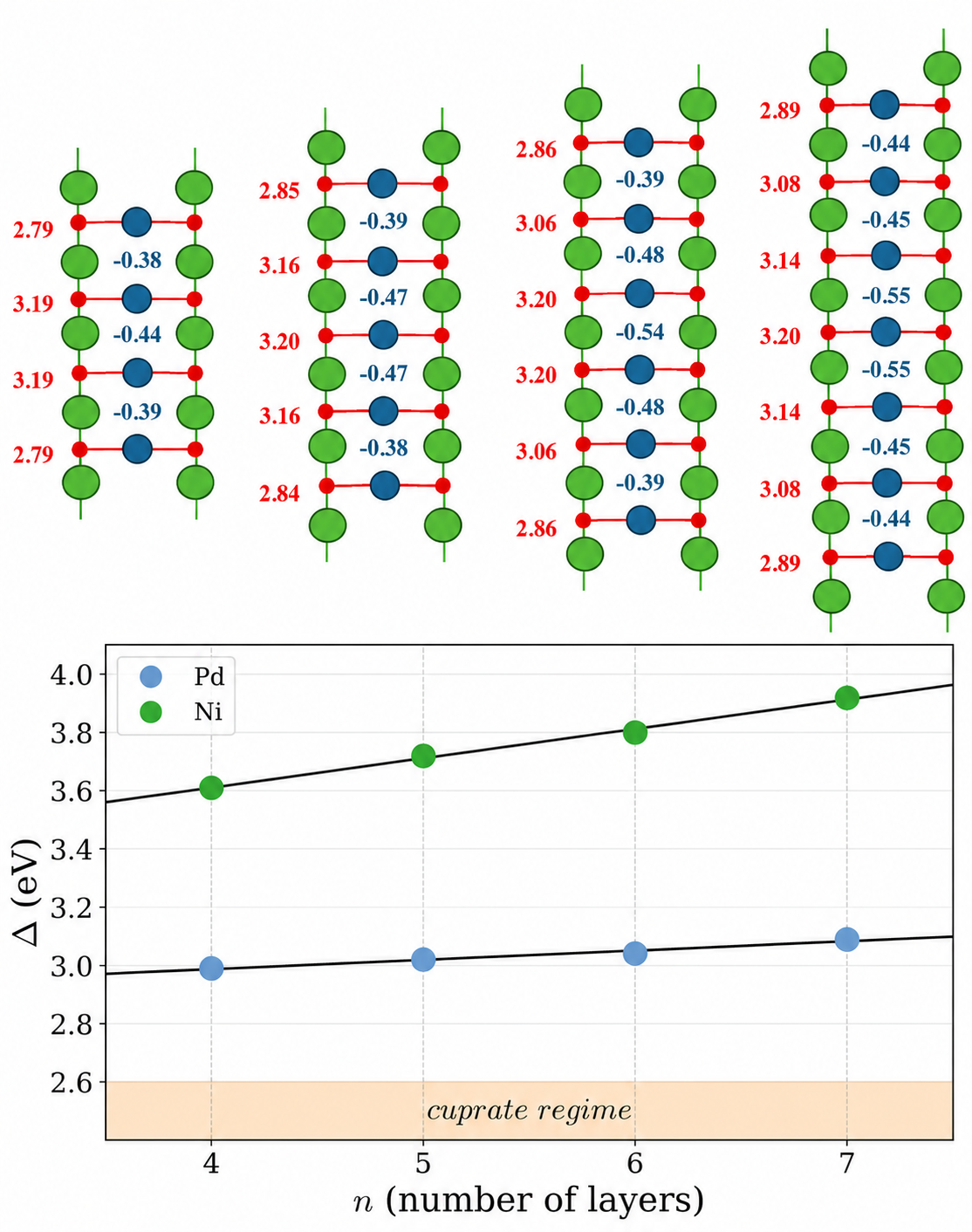}
    \caption{Top panel. Layer-resolved interlayer hoppings and charge transfer energies for the square planar palladates (all values are given in eV). The charge transfer energies are labeled in red, while the interlayer hoppings are labeled in blue. Bottom panel. Comparison of the average charge transfer energies ($\Delta$) of the square-planar nickelates (green dots) and analog palladates (blue dots). We use a maximum $\Delta$ value for cuprates of 2.6 eV, as calculated in Ref. \cite{Weber2012-hh}.} \label{ct}
\end{figure}

The atom-resolved DOSs shown in Fig. \ref{fig3} provide further confirmation of the trends seen in the band structures as the La-$d$ states shift up to higher energies in the palladates when compared to the nickelates. In addition, it is obvious from the DOS that there is a much higher degree of $p-d$ overlap in the palladates than in the corresponding nickelates. This change in $p-d$ hybridization can be quantified using the on-site energy difference between the Pd/Ni-$d_{x^2-y^2}$ and the O-$p_\sigma$ orbitals as obtained from maximally localized Wannier functions. Fig. \ref{ct} presents the average of the charge transfer energy $\Delta$ among NiO$_2$/PdO$_2$ planes, as well as the values derived for each plane in the palladates. Clear layer-by-layer modulation effects can be observed in the palladate $\Delta$ values with larger charge transfer energies being obtained in the inner planes with respect to the outer ones ($\sim$ 2.8 eV in the outer planes, $\sim$ 3.1 eV in the inner ones). Modulation effects in $\Delta$ can also be seen in the analog nickelates, as already reported in the literature \cite{LaBollita2021}. These modulation effects extend to the interlayer hoppings that are also smaller in the outer planes when compared to the inner ones (and increase by $\sim$0.1 eV for larger $n$). Importantly, both the dominant interlayer and intralayer hoppings are much larger in the square-planar palladates than in the nickelates, with the dominant $t_{pd}$ intralayer hopping being $\sim$ 1.6 eV on average in the palladates and 1.2 eV in the nickelates; reflecting also on the larger degree of $p-d$ hybridization in the Pd compounds described above. The dominant out-of-plane hopping via the $d_{z^2}$ orbitals is $\sim$ 0.2 eV in the nickelates and 0.4 eV in the palladates. This stronger interaction can be tied to the smaller $c$-lattice constant of the palladates described above.  Overall, our observed lattice modulation seems to be accompanied by layer-dependent electronic properties, similar to the behavior of multi-layer cuprates \cite{Luo2023}.

The bottom panel of Fig. \ref{ct} shows the average values of the charge-transfer energy for square-planar palladates and nickelates, where it can be observed how adding layers (i.e. going towards the infinite-layer limit) increases the charge transfer energy in both families, but much more clearly so in the nickelates. The palladates consistently have lower charge transfer energy than the nickelates ($\sim$ 3 eV) given their larger degree of $p-d$ hybridization. Indeed, the family of layered palladates falls much closer to the cuprate regime in terms of the charge-transfer energy than the analog nickelates.

\section{Magnetic Electronic Structure} \label{eStruct}

Strong antiferromagnetic correlations are considered a
key ingredient in cuprates \cite{Keimer_2015}. Using spin-polarized calculations, we have calculated and evaluated the energetics of a ferromagnetic (FM), G-type antiferromagnetic (AFM-G), and C-type antiferromagnetic (AFM-C) configuration for the $n=4-7$ square-planar palladates. For the layered nickelates, equivalent calculations can be found in Ref. \cite{LaBollita2021}. We note that a $U$ $\sim$ 3 eV is required to stabilize any of these magnetic states in the square-planar palladates. Table \ref{table2} shows the energy differences for a $U$= 4 eV. An AFM-G configuration is the magnetic ground state for all $n$s (closely competing with an AFM-C state). The derived Pd magnetic moment inside the MT sphere is reduced from its nominal 1$\mu_B$ value ($\sim$ $0.5\mu_B$) in these $n=4-7$ layered palladates due to hybridizations and also to the self-doping effect from La-$d$ states. 

\begin{table}[htbp]
    \centering
    \caption{Energy difference between a  G-type antiferromagnetic (AFM-G) state (ground state at all $n$s) and a C-type antiferromagnetic (AFM-C) as well as a ferromagnetic (FM) state. All values are given in meV. Note that a nonmagnetic state is also higher in energy.  
    }
    \vspace{0.25cm}
    \label{tab:Magnetic energies}
    \begin{tabular}{c@{\hskip 15pt}c@{\hskip 15pt}c@{\hskip 15pt}c}
        \hline
        \hline
        \textit{} & $\Delta$E$_{FM-AFM-G}$ & $\Delta$E$_{AFM-C-AFM-G}$ \\
        \hline
        \textit{$n=4$} &  6.5 & 5.0\\
        \textit{$n=5$} & 22.2 & 6.0\\
        \textit{$n=6$} &  36.2 & 15.0\\
        \textit{$n=7$} &  49.4 & 26.0\\
        \hline
        \hline
    \end{tabular}
    \label{table2}
\end{table}

\begin{figure}
    \includegraphics[width=\columnwidth]{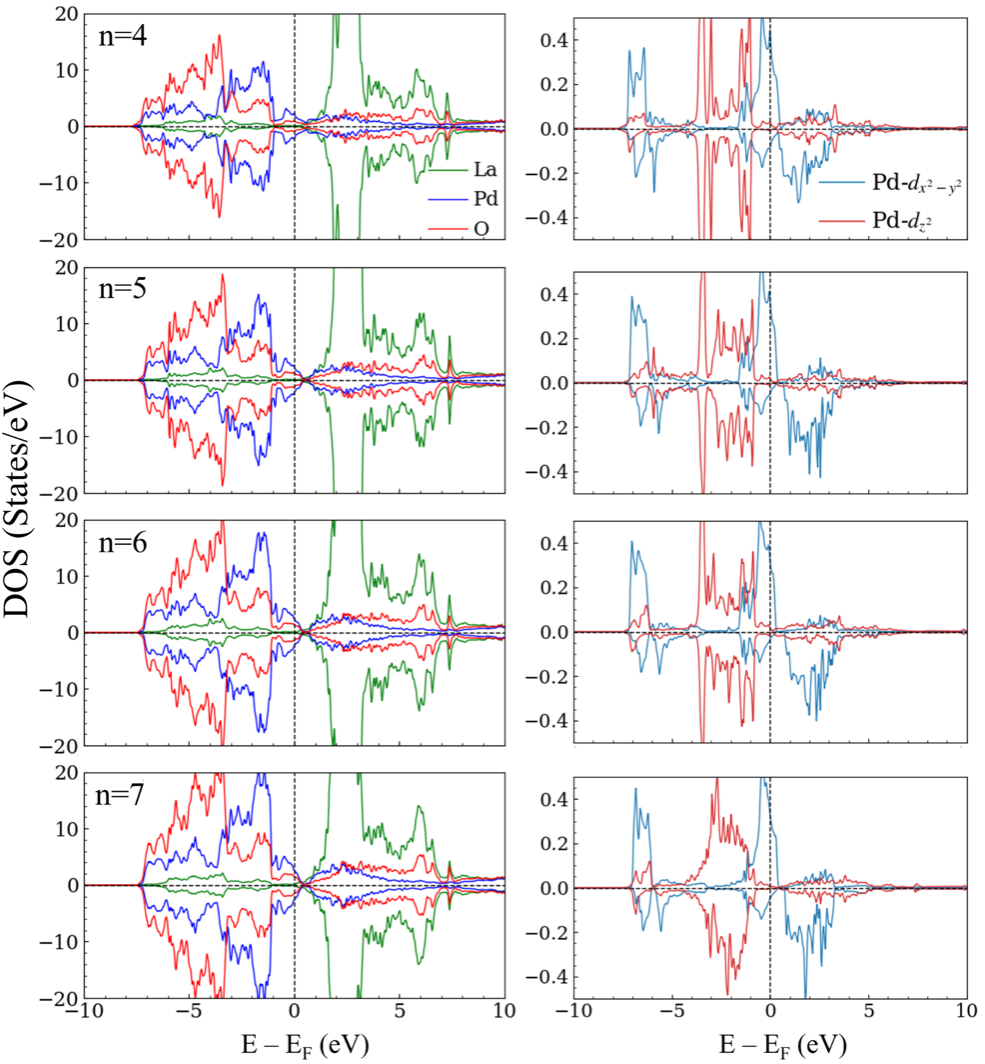}
    \caption{Atom-resolved DOS (left column) and orbital-resolved Pd-$e_g$ DOS (right column) for the $n= 4-7$ layered palladates in the AFM-G ground state configuration. The different rows correspond to $n= 4-7$ (from top to bottom).} \label{afmg}
\end{figure}

The corresponding PDOS evolution with the number of layers is shown in Fig. \ref{afmg}. The basic trends with $n$ for Pd, La, and O DOS in the AFM-G configuration follow those of the non-magnetic DOS described above in Fig. \ref{fig3}. The La-$d$ states do not contribute significantly to the low-energy physics of these palladates and slightly shift down in energy with $n$. Primarily, Pd states are present at the Fermi level and there is significant overlap between the Pd-$d$ and O-$p$ states below the Fermi level. Importantly, when looking at the orbital-resolved DOS, it is clear that the major contribution at the Fermi level comes from the Pd-$d_{x^2-y^2}$ orbital, with minimal involvement from the Pd-$d_{z^2}$ states. In the analog nickelates, there is much greater involvement from the $d_{z^2}$ states at the Fermi level and energetic competition from a FM state \cite{LaBollita2021}.

Overall, the family of layered palladates falls much closer to the cuprate regime than the analog nickelates. Their lower charge-transfer energy, together with their more single-band physics, and antiferromagnetic ground state place higher-order square-planar palladates in an intermediate regime between nickelates and cuprates and make them an interesting platform to pursue to be able to pinpoint the cuprate characteristics necessary for superconductivity.

\section{Conclusions}

We have used first-principles calculations to analyze the electronic structure of reduced higher-order square-planar palladates La$_{n+1}$Pd$_n$O$_{2n+2}$ ($n=4-7$) and compare them to their nickelate analogs. Our calculations show that the layered palladates exhibit stronger $p-d$ hybridization and hence have lower charge-transfer energies than their nickelate counterparts. In addition, they are closer to a single-$d_{x^2-y^2}$ band picture as the La-$d$ states have less interference in their low-energy physics, and they all display antiferromagnetic ground states. Hence, layered palladates exhibit more cuprate-like behavior than their corresponding nickelate analogs and represent an interesting platform to pursue in the context of unconventional superconductivity. While this family of materials has not yet been realized, given the further stability of Pd$^{1+}$ in contrast to Ni$^{1+}$, it may be possible to synthesize the members of this series directly in square-planar format-- avoiding the complications associated to topotactic reduction in the nickelates.

\section{Acknowledgments}

We acknowledge NSF Grant No. DMR-2323971 and the ASU Research Computing Center for high-performance computing resources.

\bibliography{references}

\clearpage

\end{document}